\documentclass[conference]{IEEEtran}
\IEEEoverridecommandlockouts
\usepackage{booktabs}
\usepackage{amssymb}
\usepackage{pifont}
\usepackage{multirow}
\usepackage{cite}
\usepackage{amsmath,amssymb,amsfonts}
\usepackage{algorithmic}
\usepackage{graphicx}
\usepackage{textcomp}
\usepackage{xcolor}
\usepackage{cuted}
\usepackage{float}
\usepackage{hyperref}
\usepackage{algorithm,algorithmic}
\def\BibTeX{{\rm B\kern-.05em{\sc i\kern-.025em b}\kern-.08em
    T\kern-.1667em\lower.7ex\hbox{E}\kern-.125emX}}
\begin{document}

\title{REACT-KD: Region-Aware Cross-modal Topological Knowledge Distillation for Interpretable Medical Image Classification\\

}

\author{
\IEEEauthorblockN{
Hongzhao Chen\textsuperscript{*1}
\thanks{*Co-first author},
Hexiao Ding\textsuperscript{*1},
Yufeng Jiang\textsuperscript{*1},
Jing Lan\textsuperscript{1},
Ka Chun Li\textsuperscript{1,}\textsuperscript{2},
Gerald W.Y. Cheng\textsuperscript{1}\\
Nga-Chun Ng\textsuperscript{1,}\textsuperscript{2},
Yao Pu\textsuperscript{1},
Jing Cai\textsuperscript{1},
Liang-ting Lin\textsuperscript{1},
Jung Sun Yoo\textsuperscript{\#1}\thanks{\#Corresponding author}
}
\IEEEauthorblockA{
\textsuperscript{1}Department of Health Technology and Informatics, Hong Kong Polytechnic University, Hong Kong SAR, China\\
\textsuperscript{2}Department of Nuclear Medicine and PET, Hong Kong Sanatorium and Hospital, Hong Kong SAR, China\\
Emails: \{hongzhao.chen, hexiao.ding, yufeng.jiang, jing-hti.lan, ka-chun-george.li, allen-yao.pu\}@connect.polyu.hk\\
\{wai-yeung.cheng, jing.cai, ltlin, jungsun.yoo\}@polyu.edu.hk\\
\{sam.nc.ng\}@hksh.com\\
}
}

\maketitle

\begin{abstract}
Reliable and interpretable tumor classification from clinical imaging remains a core challenge. The main difficulties arise from heterogeneous modality quality, limited annotations, and the absence of structured anatomical guidance. We present \textbf{REACT-KD}, a Region-Aware Cross-modal Topological Knowledge Distillation framework that transfers supervision from high-fidelity multi-modal sources into a lightweight CT-based student model. The framework employs a dual teacher design. One branch captures structure-function relationships through dual-tracer PET/CT, while the other models dose-aware features using synthetically degraded low-dose CT. These branches jointly guide the student model through two complementary objectives. The first achieves semantic alignment through logits distillation, and the second models anatomical topology through region graph distillation. A shared CBAM3D module ensures consistent attention across modalities. To improve reliability in deployment, REACT-KD introduces modality dropout during training, which enables robust inference under partial or noisy inputs. As a case study, we applied REACT-KD to hepatocellular carcinoma staging. The framework achieved an average AUC of 93.5\% on an internal PET/CT cohort and maintained 76.6\% to 81.5\% AUC across varying levels of dose degradation in external CT testing. Decision curve analysis further shows that REACT-KD consistently provides the highest net clinical benefit across all thresholds, confirming its value in real-world diagnostic practice. Code is available at: \url{https://github.com/Kinetics-JOJO/REACT-KD}.
\end{abstract}


\section{Introduction}
Disease classification, along with lesion segmentation and prognostic modeling, has been greatly advanced by recent developments in deep learning applied to computed tomography (CT) imaging ~\cite{zhou2021review}. While models trained on high-resolution or denoised low-dose CT (LDCT) scans have demonstrated strong performance~\cite{rayed2024segmentation,chen2024lowdose}, real-world deployment remains limited due to a mismatch between curated datasets and the variability of clinical imaging conditions. One major limitation arises from diverse CT acquisition protocols, particularly in radiation dose. For instance, LDCT is widely employed in lung cancer screening to minimize patient exposure, but often lacks the spatial fidelity required for diagnostic or staging purposes~\cite{zhang2024denoising}. Hepatocellular carcinoma (HCC) exemplifies this challenge, as its diagnosis typically relies on multiphase CT or functional modalities like positron emission tomography (PET)~\cite{nguyen2024advanced}. 

Although PET/CT provides complementary metabolic and anatomical cues, the CT component in PET often uses low-dose settings, resulting in increased noise and reduced resolution~\cite{chaudhari2021lowcount}. PET itself is also subject to variations in count statistics, which further complicate fusion~\cite{chaudhari2021lowcount}. Additionally, differences in spatial resolution, acquisition parameters, and anatomical alignment across modalities present major challenges for multi-modal integration~\cite{kumar2024fusion,zhang2024survey}. While multi-modal learning aims to leverage these complementary features, most approaches assume access to well-aligned, high-quality paired data, which are seldom available in clinical settings~\cite{zhou2024precision,wang2023sharedspecific,han2022dynamical}. These constraints necessitate the development of robust, interpretable frameworks capable of operating under missing or degraded modality conditions.

Knowledge distillation (KD) has emerged as an effective strategy to address these challenges~\cite{zhou2024precision}. By transferring supervision from a complex teacher model to a lightweight student, KD allows compact models to retain high performance and generalization~\cite{ahmad2024multi}. Classical KD approaches such as logits distillation~\cite{hinton2015distilling,zhao2022decoupled}, attention transfer~\cite{yang2024vitkd}, and relational learning~\cite{liu2023graphbasedknowledgedistillationsurvey} have proven successful in tasks including classification, segmentation, and cross-modal fusion~\cite{song2025lightweight,gao2025fuzzy,yu2025sleep}. Particularly, there have been recent advances in multi-teacher and cross-modal KD~\cite{wang2023learnable,li2025mst}, which facilitate robust learning from heterogeneous modalities even when some are unavailable during clinical practice.

In this study, we present a dual-teacher cross-modal distillation framework to address the limitations regarding clinically acquired imaging data and ensure reliable task performance. HCC is selected as a case study due to its diagnostic reliance on both functional and anatomical imaging. Although our work focuses on classification, the framework is extensible to other medical tasks. The proposed method, \textbf{REACT-KD}, introduces a dual-teacher paradigm: a structure–function-aware teacher trained on dual-tracer PET/CT and a dose-aware teacher trained on simulated LDCT. A lightweight student model is trained under modality dropout to ensure resilience to missing or noisy inputs. Beyond conventional logits supervision, we propose a Region Graph Distillation (RGD) module that leverages segmentation-derived anatomical graphs to enforce structure-aware feature consistency. A 3D based Convolutional Block Attention Module (CBAM3D)~\cite{Woo2018CBAM} further promotes spatially aligned feature learning across modalities. Experimental evaluation on in-house and external HCC datasets demonstrates the robustness and generalizability of REACT-KD under varying clinical situations.
\noindent \textbf{The main contributions of this work are summarized as follows:}
\begin{enumerate}
\item We propose a dual-teacher distillation framework that integrates structure–function and dose-aware supervision, enabling robust cross-modal knowledge transfer from high-fidelity PET/CT to a compact CT-only model.
\item We introduce Region Graph Distillation, a topology-aware mechanism that captures anatomical relationships from segmentation-derived graphs to guide structure-consistent student learning.
\item We present a modality-decoupled training scheme with modality dropout, improving the student’s resilience to missing or noisy inputs and supporting deployment in diverse clinical settings.
\end{enumerate}

\section{Methodology}

\subsection{Method Overview}
We propose a two-stage knowledge distillation framework comprising a dual-branch teacher and a lightweight student. The teacher is trained on multi-source volumetric inputs to learn robust representations, while the student is distilled under modality dropout to handle incomplete inputs.

\begin{figure*}[t]
    \centering
    \includegraphics[width=\textwidth]{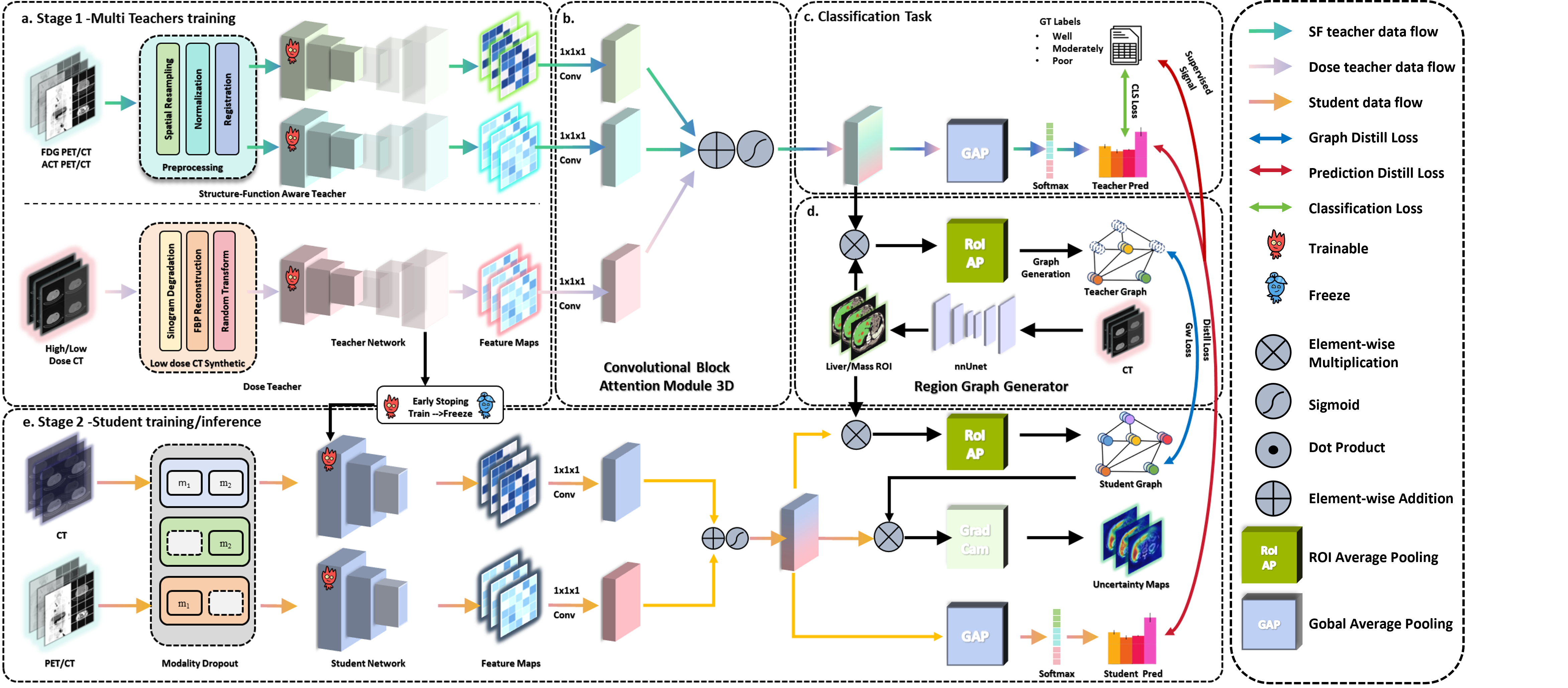}
    \caption{\textbf{Overall pipeline of our distillation framework.} 
    (a) Dual-branch teacher model: a structure-function-aware (SF) path processes dual-tracer PET/CT images from the institutional cohort, and a dose-aware path handles simulated dose-varied CT scans derived from the LiTS17 dataset. 
    (b) CT features from both paths are extracted using a shared SwinUNETR encoder, while PET features are separately encoded.
    (c) Feature fusion is performed via a shared CBAM3D module to enhance discriminative regions.
    (d) Region graphs are constructed based on liver and tumor masks for topological supervision.
    (e) A student network learns under modality dropout with guidance from logits-based and region graph distillation losses.}
    \label{fig:pipeline}
\end{figure*}

The framework introduces two complementary distillation objectives:  
(i) \textit{Logits Distillation} transfers soft class-level predictions to enhance output alignment;  
(ii) \textit{Region Graph Distillation} encodes anatomical topology from liver–tumor masks to guide structure-aware learning.  
Together, these objectives improve generalization and interpretability. The overall pipeline is shown in Fig.~\ref{fig:pipeline}.

\subsection{Logits Distillation}
Logits distillation is a classical knowledge distillation technique originally introduced by Hinton et al.~\cite{hinton2015distilling}. In this approach, the student network is trained to approximate the soft class probability distribution produced by the teacher network, thereby capturing not only the hard labels but also the relative class-wise confidences learned by the teacher.

Let $\mathbf{z}^{(T)} \in \mathbb{R}^C$ and $\mathbf{z}^{(S)} \in \mathbb{R}^C$ denote the logits (pre-softmax outputs) of the teacher and student networks, respectively, where $C$ is the number of target classes. The soft predictions are obtained using a temperature-scaled softmax function:

\begin{equation} \sigma(\mathbf{z}, \tau)i = \frac{\exp(z_i / \tau)}{\sum{j=1}^{C} \exp(z_j / \tau)}, \end{equation}

where $\tau$ is the temperature parameter that controls the smoothness of the output distribution. The logits distillation loss is defined as the Kullback–Leibler (KL) divergence between the softened outputs:

\begin{equation} \mathcal{L}_{\text{logit}} = \text{KL} \left( \sigma\left(\mathbf{z}^{(T)}, \tau \right) ;||; \sigma\left(\mathbf{z}^{(S)}, \tau \right) \right). \end{equation}

By minimizing $\mathcal{L}_{\text{logit}}$, the student network is encouraged to replicate the teacher’s predictive behavior, thereby learning richer inter-class relationships and enhancing generalization performance.

\subsection{Topological Distillation Via Lesion Region}

Topological distillation via lesion region leverages graph-based techniques to facilitate relational knowledge transfer by modeling semantic and spatial dependencies in medical imaging. Zhao et al.~\cite{mskd} introduced the MSKD framework, which constructs region-level graphs from uniform 2D patches extracted from intermediate feature maps. Patch-level features are computed via global average pooling (GAP), and inter-patch relationships are captured through cosine similarity. Despite its effectiveness, this method exhibits two key limitations, (1) uniform patch partitioning results in dense but semantically weak graphs and introduces redundancy; (2) pairwise similarity may fail to capture complex contextual dependencies in clinical imaging.

To address these issues, we propose a 3D, region-of-interest (ROI)-aware graph distillation framework. Instead of uniform patches, we define anatomically meaningful ROIs from segmentation masks, including the liver parenchyma and one or more tumor masses. Each ROI is encoded as a graph node, enabling semantically interpretable and structurally consistent representation learning. This approach enhances both spatial fidelity and anatomical alignment during the distillation process.

Let $\mathbf{F}_s, \mathbf{F}_t \in \mathbb{R}^{C \times D \times H \times W}$ denote the volumetric features extracted from the student and teacher networks, respectively. Given ROI masks $\mathcal{M}_k$, we define each node feature $\mathbf{r}_k \in \mathbb{R}^C$ using masked global average pooling: 

\begin{equation}
\mathbf{r}_k = \frac{1}{|\Omega_k|} \sum_{\mathbf{x} \in \Omega_k} \mathbf{F}(\mathbf{x}), \quad \Omega_k = \left\{ \mathbf{x} \mid \mathcal{M}_k(\mathbf{x}) = 1 \right\},
\end{equation}

where each ROI $\Omega_k$ corresponds to either the liver region or an individual tumor mass. This formulation naturally accommodates varying lesion counts and spatial distributions.

Importantly, while tumor masses are anatomically located within the liver, we explicitly represent each tumor and the liver as separate nodes to preserve their semantic and structural individuality. This modeling choice reflects clinically significant characteristics such as tumor multiplicity and spatial dispersion. From a structural perspective, these characteristics correspond to topological features, such as the number of disconnected components within the organ. In this context, tumor multiplicity can be interpreted as a first-order topological property, which is relevant to the diagnosis and prognosis of HCC.

We construct the region graph $\mathcal{G} = (\mathcal{V}, \mathcal{E})$ by treating each region feature $\mathbf{r}k$ as a node and defining edge weights via cosine similarity: \begin{equation} S{p,q} = \frac{ \langle \mathbf{r}_p, \mathbf{r}_q \rangle }{ | \mathbf{r}_p |_2 \cdot | \mathbf{r}_q |_2 }. \end{equation}

To transfer graph-level knowledge from the teacher to the student network, we employ two conventional loss terms: the node-wise feature alignment loss and the edge-wise similarity alignment loss: 
\begin{align}
\mathcal{L}_{\text{node}} &= \frac{1}{N} \sum_{k=1}^{N} \left\| \frac{\mathbf{r}^s_k}{\|\mathbf{r}^s_k\|_2} - \frac{\mathbf{r}^t_k}{\|\mathbf{r}^t_k\|_2} \right\|_2^2, \\
\mathcal{L}_{\text{edge}} &= \frac{1}{N^2} \sum_{i,j=1}^{N} \left( S^s_{i,j} - S^t_{i,j} \right)^2.
\end{align}

However, due to inter-patient variability in lesion burden and tumor distribution, the number of ROI nodes may vary across cases, resulting in region graphs with mismatched cardinality and topology. In such settings, enforcing rigid one-to-one node or edge correspondence becomes impractical. To overcome this, we introduce a Gromov-Wasserstein (GW) discrepancy loss~\cite{xu2019gromov} that aligns two graphs by jointly considering node feature similarity and pairwise structural relationships: \begin{equation} \mathcal{L}_{\text{GW}} = \min{\pi \in \Pi(\mu, \nu)} \sum_{i,j,k,l} \left| S^s_{i,j} - S^t_{k,l} \right|^2 \pi_{i,k} \pi_{j,l}, \end{equation} where $\pi \in \mathbb{R}_+^{N_s \times N_t}$ represents a soft transport plan between student and teacher nodes, and $\Pi(\mu, \nu)$ denotes the set of admissible transport plans with marginal distributions $\mu$ and $\nu$, which are typically assumed to be uniform in practice.

This formulation enables flexible alignment of graph structures with differing node counts and spatial configurations, thereby preserving topological consistency across heterogeneous cases. The final region graph distillation objective combines all three loss terms: \begin{equation} \mathcal{L}_{\text{RGD}} = \lambda_1 \mathcal{L}_{\text{node}} + \lambda_2 \mathcal{L}_{\text{edge}} + \lambda_3 \mathcal{L}_{\text{GW}}, \end{equation} where $\lambda_1$, $\lambda_2$, and $\lambda_3$ are weighting coefficients for node-level, pairwise, and global structural alignment, respectively. Empirical results demonstrate that incorporating the GW term enhances the student network’s capacity to replicate spatial and structural patterns learned by the teacher, particularly in cases with multiple lesions or irregular topologies.

\subsection{Teacher Network}

The teacher network employs a dual-branch architecture with modality-specific encoders to extract rich volumetric features. Each branch is built upon the SwinUNETR backbone~\cite{hatamizadeh2022swin}, due to its capability to process 3D medical images with a hierarchical Transformer architecture.

One branch processes dual-tracer PET/CT inputs to capture complementary structural and functional cues, while the other handles dose-varied CT scans to enhance robustness under acquisition variability. Both encoders share the same SwinUNETR design and jointly provide supervision signals during distillation.

All encoders used in the teacher branches follow the SwinUNETR architecture. The input 3D volume is partitioned into non-overlapping patches of size $P\times M\times M$, which are linearly embedded into a sequence of tokens. Each token sequence $X \in \mathbb{R}^{N_w \times C}$ is projected to query ($Q$), key ($K$), and value ($V$) matrices:
\begin{equation}
Q = XW_Q, \quad K = XW_K, \quad V = XW_V,
\end{equation}
where $W_Q, W_K, W_V \in \mathbb{R}^{C \times d}$ are learnable weights. Within each 3D window, attention is computed as:
\begin{equation}
\text{Attention}(Q, K, V) = \text{Softmax}\left(\frac{QK^\top}{\sqrt{d}} + B\right)V,
\end{equation}
where $B \in \mathbb{R}^{N_w \times N_w}$ represents the relative position bias. Alternating window-based multi-head self-attention (W-MSA) and shifted-window multi-head self-attention (SW-MSA) blocks are applied to model local and non-local context. SW-MSA introduces a cyclic shift by 
$\left(\left\lfloor \frac{P}{2} \right\rfloor, \left\lfloor \frac{M}{2} \right\rfloor, \left\lfloor \frac{M}{2} \right\rfloor \right)$
voxels and applies masking to restrict attention computation within valid shifted windows. Each block is updated as:
\begin{equation}
\begin{aligned}
\hat{z}^l &= \text{W-MSA}(\text{LN}(z^{l-1})) + z^{l-1}, \\
z^l &= \text{MLP}(\text{LN}(\hat{z}^l)) + \hat{z}^l, \\
\hat{z}^{l+1} &= \text{SW-MSA}(\text{LN}(z^l)) + z^l, \\
z^{l+1} &= \text{MLP}(\text{LN}(\hat{z}^{l+1})) + \hat{z}^{l+1},
\end{aligned}
\end{equation}
where W-MSA and SW-MSA denote regular and shifted window multi-head self-attention mechanisms. MLP refers to a two-layer feed-forward network applied to each token independently:
\begin{equation}
\text{MLP}(x) = \text{GELU}(xW_1 + b_1)W_2 + b_2.
\end{equation}


The outputs from both branches are fused through a 3D fusion module, which enhances discriminative feature representations. This module, detailed in the follow section, produces refined embeddings for tumor grade prediction and guides the student network through region-graph supervision.

\subsection{Student Network}

The student network is constructed using a SegResNet-based encoder~\cite{myronenko20183dmribraintumor}, with this architecture adopted due to its lightweight design and comprehensive integration within the MONAI framework. It comprises residual convolutional blocks consisting of instance normalization, ReLU activations, and stacked 3D convolutions. Given an input $F^{(l)}$ at layer $l$, the residual unit is defined as:
\begin{align}
Z^{(l)} &= \text{ReLU}(\text{IN}(\text{Conv}_{3D}(F^{(l)}))), \\
F^{(l+1)} &= F^{(l)} + \text{Conv}_{3D}(Z^{(l)}).
\end{align}

Downsampling is performed using strided 3D convolutions, while $1\times1\times1$ kernels are used to reduce channel dimensions:
\begin{align}
F^{(l+1)}_{\text{down}} &= \text{Conv}_{3D}^{(s=2)}(F^{(l+1)}), \\
F_{\text{reduced}} &= \text{Conv}_{1\times1\times1}(F).
\end{align}

To improve robustness under missing or noisy input conditions, a modality dropout strategy is applied during training, where either the PET or CT input is randomly masked with a predefined probability $p_{\text{drop}}$. This strategy encourages generalization across heterogeneous input settings. This strategy promotes generalization across heterogeneous input settings, while the 3D attention module is shared with the teacher to ensure consistent saliency modeling.


\subsection{Attention-Guided Feature Fusion}

To improve volumetric feature discrimination under heterogeneous modality inputs, we adopt a 3D extension of the Convolutional Block Attention Module (CBAM3D)~\cite{Woo2018CBAM}. This module sequentially applies channel and spatial attention to adaptively recalibrate feature responses, highlighting semantically relevant anatomical regions.

To ensure consistent attention behavior, CBAM3D is shared between the teacher and student networks. Rather than performing direct feature-level distillation, which may be sensitive to modality dropout, CBAM3D serves as an implicit fusion mechanism to harmonize cross-modal attention. This facilitates robust and interpretable representation learning without the need for explicit feature alignment.

\subsection{Training Strategy}
We adopt a two-stage training strategy to effectively transfer semantic and structural knowledge from the teacher to the student network while enhancing robustness under incomplete modality conditions.

\paragraph{\textbf{Stage 1} Teacher Supervision} 
The teacher network is trained using full dual-modality inputs, with supervision provided by the focal loss for tumor grade classification:
\begin{equation}
\mathcal{L}_{\text{focal}} = -\sum_{c} \alpha_c (1 - p_c)^\gamma y_c \log(p_c),
\end{equation}
where $p_c$ is the predicted probability for class $c$, $y_c$ is the one-hot encoded label, $\alpha_c$ is the class-balancing factor, and $\gamma$ is the focusing parameter. Region graphs are constructed during this phase using expert-verified liver and tumor masks, though no structural loss is applied at this stage.

\paragraph{\textbf{Stage 2} Student Distillation} 
The student network is trained under a modality dropout mechanism, in which either PET or CT is randomly omitted to simulate real-world incomplete input scenarios. The student is supervised by the same focal loss $\mathcal{L}_{\text{focal}}$ used during teacher training, in conjunction with the distillation objectives previously defined. The overall training objective is given by:
\begin{equation}
\mathcal{L}_{\text{total}} = \lambda_1 \mathcal{L}_{\text{focal}} + \lambda_2 \mathcal{L}_{\text{logits}} + \lambda_3 \mathcal{L}_{\text{RGD}}.
\end{equation}
where $\lambda_1$, $\lambda_2$, and $\lambda_3$ are balancing weights tuned via cross-validation. This strategy ensures that the student acquires both classification capabilities and graph-structured anatomical awareness, enabling it to perform robustly across heterogeneous and potentially incomplete clinical inputs.

\section{experiment}

\subsection{Data Description}

We constructed a multi-source dataset to support supervised training, representation learning, and external validation for HCC tumor grade classification. The dataset includes one institutional PET/CT cohort and two public CT datasets with varying modalities and clinical scenarios:

\begin{itemize}
    \item \textbf{Hospital HCC Cohort (Private)}: 194 dual-tracer PET/CT scan pairs from 97 HCC patients, each imaged with both $^{18}$F-FDG and $^{11}$C-Acetate. Every scan pair includes co-registered PET and CT volumes with dimensions of $512 \times 512 \times 148$. Tumor grades are labeled based on the Edmondson-Steiner system (Grade 1-well differentiated; Grade 2-moderately differentiated; Grade 3-poorly differentiated). Associated clinical metadata includes age, sex, hepatitis status and alpha-fetoprotein.

    \item \textbf{LiTS17 Benchmark (Public)}: 131 contrast-enhanced CT volumes with liver and tumor masks but no grade labels, from the LiTS challenge~\cite{bilic2023lits}. Scans exhibit variable resolution (42–1026 slices; 0.56–1.0 mm in-plane spacing).
    
    \item \textbf{HCC-TACE-Seg (Public)}: 105 multiphasic contrast-enhanced CT scans with expert segmentation and histopathological labels~\cite{Moawad2023HCC_TACESeg}. Data reflect real-world TACE treatment cases under clinical imaging protocols.
\end{itemize}

\subsection{Data Preprocessing}

CT volumes were clipped to $[-160, 240]$ HU, globally z-score normalized, and resampled to $64 \times 224 \times 224$ using linear interpolation. PET volumes were processed by clipping to the 1st–99th percentile range, min-max normalized to $[0, 1]$, and standardized per volume. PET resampling followed the spatial geometry of paired CTs to ensure alignment.

Segmentation masks were refined through the extraction of the largest connected component, followed by two iterations of binary morphological closing and the application of nearest-neighbor interpolation. Each case was processed to generate three input representations, comprising full normalized volumes, intensity-masked maps, and raw-value lesion crop masks.







\subsection{Implementation Details}

All models were implemented in PyTorch and trained on dual NVIDIA RTX 4090 GPUs using 3-fold cross-validation. The teacher network was trained for 150 epochs (batch size 2, learning rate $2\times10^{-4}$ with cosine decay), and the student for 50 epochs (batch size 8, learning rate $1\times10^{-4}$) under modality dropout. The total loss combined focal loss (with class weights 1.5/1.0/2.0), logits distillation (temperature $\tau=4$), and region graph-based Gromov-Wasserstein loss ($\lambda_{\text{GW}}=2.0$). During training, 3D data augmentations including affine transforms and elastic deformations were applied.

\subsection{Model Evaluations}

We evaluate the proposed framework through internal cross-validation and external generalization testing, reflecting diverse clinical conditions.

\subsubsection{Internal Validation}
Three-fold patient-level cross-validation was conducted on the institutional PET/CT cohort. Dual-teacher models were trained on full dual-modality inputs, while the student network was optimized under modality dropout. Performance was reported on held-out patients in each fold. Ablation studies assessed the contributions of each distillation loss, the CBAM3D module, and robustness under missing modality settings.

\subsubsection{External Validation}

External generalization was tested on the HCC-TACE-Seg dataset using contrast-enhanced CT scans with differing acquisition protocols. The student model, trained solely on internal data, was evaluated without fine-tuning. During teacher training, LiTS17 CT volumes (without labels) were included to enhance dose-aware representation learning, improving adaptability to variable imaging quality.

\subsubsection{Evaluation Metrics}

Performance on the three-class tumor grade classification task was assessed using macro-averaged accuracy (ACC), precision, recall, F1-score, and area under the receiver operating characteristic curve (AUC). Metric definitions are as follows:

\begin{equation}
\text{Accuracy} = \frac{TP + TN}{TP + TN + FP + FN}, \tag{19}
\end{equation}
\begin{equation}
\text{Precision}_c = \frac{TP_c}{TP_c + FP_c}, \tag{20}
\end{equation}
\begin{equation}
\text{Recall}_c = \frac{TP_c}{TP_c + FN_c}, \tag{21}
\end{equation}
\begin{equation}
\text{F1}_c = \frac{2 \times \text{Precision}_c \times \text{Recall}_c}{\text{Precision}_c + \text{Recall}_c}, \tag{22}
\end{equation}
\begin{equation}
\text{AUC}_c = \int_{0}^{1} \text{TPR}_c(t) \, d\text{FPR}_c(t). \tag{23}
\end{equation}

Here, \(TP\), \(TN\), \(FP\), and \(FN\) denote true positives, true negatives, false positives, and false negatives. Subscript \(c\) indicates class-wise calculation for the three tumor grades (well, moderate, poor). The AUC was computed using a one-vs-rest strategy and averaged across classes.

\section{Results}

This section presents a comprehensive empirical evaluation of the proposed dual-teacher distillation framework. We assess its effectiveness from multiple perspectives, including comparative performance against existing distillation baselines, contributions of architectural components, robustness under varying input conditions, and interpretability of the learned representations. The results collectively demonstrate the advantages of incorporating modality-specific supervision and topological guidance into the distillation process.

\subsection{Comparison with Existing Distillation Strategies}
We compare our framework against six representative knowledge distillation methods applicable to volumetric, multi-modal classification: Logits Distillation~\cite{hinton2015distilling}, Attention Transfer (AT)~\cite{zagoruyko2016paying}, CIRKD~\cite{cirkd}, MSKD~\cite{mskd}, MTCM-KD~\cite{mtcmkd}, and AWMTKD~\cite{eddardaa2025adaptive}. These baselines were implemented with matched teacher–student architectures and support coarse- or token-level supervision. Other fine-grained methods such as CWD~\cite{shu2021channelwise}, CRD~\cite{crd}, and IFVD~\cite{ifvd} require spatial alignment and architectural homogeneity, making them incompatible with our dual-branch, modality-specific design.

As summarized in Table~\ref{tab:distill_compare}, our method outperforms all baselines, achieving 79.4\% F1-score, 93.5\% AUC, and 79.9\% accuracy. Compared to Logits Distillation, it yields +10.0\% in F1 and +9.7\% in AUC, indicating improved class separability and calibration. While attention-based methods (AT, CIRKD) provide limited gains, and relational methods (MSKD, MTCM-KD) improve AUC under homogeneous setups, they fall short under heterogeneous, multi-source conditions. AWMTKD benefits from token-level fusion but lacks anatomical guidance. In contrast, our dual-teacher approach integrates modality-aware supervision and topological regularization, enabling consistent and interpretable improvements.

\begin{table*}[t]
\centering
\caption{
Comparison with representative knowledge distillation methods. ``\checkmark'' denotes support; ``\ding{55}'' denotes lack of support. All classification metrics are reported as macro-averaged values (percentage format) with standard deviation over three-fold cross-validation.
}

\begin{tabular}{lccc|ccccc}
\toprule
\multirow{2}{*}{\textbf{Method}} &
\multicolumn{3}{c|}{\textbf{Distillation Type}} &
\multicolumn{5}{c}{\textbf{Prediction Performance (\% Mean $\pm$ STD)}} \\
\cmidrule(r){2-4} \cmidrule(l){5-9}
& \textbf{Logits} & \textbf{Feature} & \textbf{Graph}
& \textbf{Accuracy} & \textbf{Precision} & \textbf{Recall} & \textbf{F1 Score} & \textbf{AUC} \\
\midrule
KD~\cite{hinton2015distilling}           & \checkmark & \ding{55} & \ding{55} & 70.1 $\pm$ 4.2 & 69.1 $\pm$ 4.2 & 72.2 $\pm$ 4.0 & 69.4 $\pm$ 4.8 & 83.8 $\pm$ 2.8 \\
AT~\cite{zagoruyko2016paying}            & \ding{55} & \checkmark & \ding{55} & 72.2 $\pm$ 3.3 & 71.4 $\pm$ 2.9 & 73.9 $\pm$ 5.7 & 71.7 $\pm$ 3.6 & 84.3 $\pm$ 2.7 \\
CIRKD~\cite{cirkd}                       & \checkmark & \ding{55} & \checkmark & 74.7 $\pm$ 4.2 & 72.4 $\pm$ 4.2 & 75.5 $\pm$ 5.4 & 73.2 $\pm$ 5.0 & 84.8 $\pm$ 2.4 \\
MSKD~\cite{mskd}                         & \checkmark & \ding{55} & \checkmark & 74.1 $\pm$ 6.4 & 71.9 $\pm$ 6.9 & 73.9 $\pm$ 5.8 & 72.2 $\pm$ 6.9 & 90.7 $\pm$ 3.8 \\
MTCM-KD~\cite{mtcmkd}                    & \ding{55} & \checkmark & \ding{55} & 76.5 $\pm$ 6.8 & 75.9 $\pm$ 6.6 & 77.5 $\pm$ 9.4 & 76.4 $\pm$ 7.5 & 92.1 $\pm$ 4.5 \\
AWMTKD~\cite{eddardaa2025adaptive}                     & \checkmark & \checkmark & \checkmark & 78.1 $\pm$ 3.7 & 76.5 $\pm$ 3.3 & 78.1 $\pm$ 3.8 & 77.1 $\pm$ 3.3 & 93.4 $\pm$ 2.2 \\
\textbf{Ours}                            & \checkmark & \ding{55} & \checkmark & \textbf{79.9 $\pm$ 6.2} & \textbf{78.7 $\pm$ 6.5} & \textbf{81.8 $\pm$ 5.3} & \textbf{79.4 $\pm$ 6.7} & \textbf{93.5 $\pm$ 2.2} \\
\bottomrule
\end{tabular}
\label{tab:distill_compare}
\end{table*}

\subsection{Ablation Analysis}
To systematically assess the role of each component in our framework, we conduct a series of ablation experiments under internal cross-validation:

\begin{itemize}
    \item \textbf{w/o Dose Teacher}: Removes the auxiliary LDCT teacher to assess the benefit of dual-teacher supervision.
    \item \textbf{w/o RGD}: Disables the graph-based structural loss, training the student with logits-only supervision.
    \item \textbf{w/o CBAM3D}: Removes the CBAM3D module in both teacher and student networks to evaluate the impact of attention-based feature enhancement.
    \item \textbf{Full Model}: The proposed configuration incorporating all components.
\end{itemize}

Table~\ref{tab:ablation_full} summarizes the results. Excluding the dose-aware teacher leads to a 3.3\% drop in AUC, indicating that modality-adaptive supervision improves robustness under varying acquisition conditions. Removing the RGD causes the largest performance drop, with F1-score reduced by 5.5\% and AUC by 6.3\%, confirming the importance of spatial structure modeling through anatomical topology. Disabling the CBAM3D module results in a 4.2\% decrease in F1, suggesting that attention-guided recalibration strengthens volumetric feature encoding. These findings highlight the complementary roles of modality supervision, topological regularization, and attention-based enhancement in improving multi-class tumor grade classification.

\begin{table}[ht]
\centering
\caption{Combinatorial ablation study of core components under internal validation. ``\checkmark'' denotes presence; ``\ding{55}'' denotes removal. Metrics are reported as mean ± std over three-fold cross-validation.}
\begin{tabular}{ccc|ccc}
\toprule
\multicolumn{3}{c|}{\textbf{Module}} & \multicolumn{3}{c}{\textbf{Prediction Metrics (Mean\%\ ± STD)}} \\
\cmidrule(lr){1-3} \cmidrule(l){4-6}
\textbf{Dose-T} & \textbf{RGD} & \textbf{CBAM3D} & \textbf{Accuracy} & \textbf{F1 Score } & \textbf{AUC} \\
\midrule
\checkmark & \checkmark & \checkmark & \textbf{79.9 ± 6.2} & \textbf{79.4 ± 6.7} & \textbf{93.5 ± 2.2} \\
\ding{55}  & \checkmark & \checkmark & 76.1 ± 4.9 & 75.0 ± 4.1 & 90.2 ± 1.8 \\
\checkmark & \ding{55}  & \checkmark & 74.4 ± 5.1 & 73.9 ± 3.2 & 87.2 ± 2.0 \\
\checkmark & \checkmark & \ding{55}  & 77.5 ± 3.8 & 75.2 ± 4.9 & 89.2 ± 1.9 \\
\ding{55}  & \ding{55}  & \checkmark & 72.0 ± 4.2 & 72.8 ± 3.3 & 84.4 ± 2.1 \\
\ding{55}  & \checkmark & \ding{55}  & 73.5 ± 4.3 & 71.9 ± 2.5 & 86.9 ± 2.2 \\
\checkmark & \ding{55}  & \ding{55}  & 72.6 ± 2.1 & 73.0 ± 2.0 & 84.6 ± 2.0 \\
\ding{55}  & \ding{55}  & \ding{55}  & 70.1 ± 4.2 & 69.4 ± 4.2 & 83.8 ± 2.8 \\
\bottomrule
\end{tabular}
\label{tab:ablation_full}
\end{table}

\begin{table}[ht]
\centering
\caption{Internal validation under different modality inputs. Metrics are reported as mean ± std over three-fold cross-validation.}
\begin{tabular}{ccc|ccc}
\toprule
\multicolumn{3}{c|}{\textbf{Modality}} & \multicolumn{3}{c}{\textbf{Prediction Performance (Mean\%\ $\pm$ STD)}} \\
\cmidrule(lr){1-3} \cmidrule(l){4-6}
\textbf{PET} & \textbf{CT} &  & \textbf{Accuracy} & \textbf{F1 Score} & \textbf{AUC} \\
\midrule
\checkmark & \checkmark & & \textbf{79.9 ± 6.2} & \textbf{79.4 ± 6.7} & \textbf{93.5 ± 2.2} \\
\checkmark & \ding{55}  & & 71.3 ± 3.9 & 70.8 ± 2.0 & 86.2 ± 1.8 \\
\ding{55}  & \checkmark & & 74.4 ± 5.1 & 73.9 ± 3.2 & 83.3 ± 2.0 \\
\bottomrule
\end{tabular}
\label{tab:internal_dropout}
\end{table}

\begin{table}[ht]
\centering
\caption{External validation on HCC-TACE-Seg using simulated CT images under varying degradation levels. Metrics are reported as mean.}
\begin{tabular}{lccc}
\toprule
\textbf{Setting} & \textbf{Degradation Level} & \textbf{F1 Score} & \textbf{AUC} \\
\midrule
Native Dose CT       & None (Original)              & 62.80 & \textbf{81.51} \\
Mild Degradation CT     & $1 \times 10^4$              & \textbf{63.29} & 80.06 \\
Severe Degradation CT    & $5 \times 10^4$              & 54.21& 76.58 \\
Mixed Native + Severe CT & Native / $5 \times 10^4$     & 57.60 & 79.96  \\
\bottomrule
\end{tabular}
\label{tab:dose_sensitivity}
\end{table}

\subsection{Robustness and Generalization}

We evaluate robustness under partial modality inputs and assess generalization on external data. For internal validation, either PET or CT inputs were selectively removed at inference. As reported in Table~\ref{tab:internal_dropout}, the full PET-CT configuration yielded the highest performance, achieving an F1-score of 79.4\% and an AUC of 93.5\%. Removing CT led to an 8.6\% reduction in F1-score and a 7.3\% drop in AUC, underscoring the importance of anatomical information. In contrast, removing PET resulted in a 10.2\% decrease in AUC, highlighting the complementary metabolic information provided by PET. These results confirm that the model retains robust performance even when only partial modalities are available.

Generalization was further evaluated on the HCC-TACE-Seg dataset using CT scans with varying noise levels. As shown in Table~\ref{tab:dose_sensitivity}, performance was assessed under native, mild, severe, and mixed degradation conditions. Simulated noise was introduced via sinogram-domain perturbation following Zeng et al.~\cite{zeng2015simulation}. Although performance declined under severe noise, the AUC dropped by no more than 4.93\%, indicating high tolerance to image degradation. Notably, mild degradation slightly improved the F1-score, suggesting a potential regularization benefit from noise variation.

\begin{figure}[ht]
    \centering
    \includegraphics[width=\linewidth]{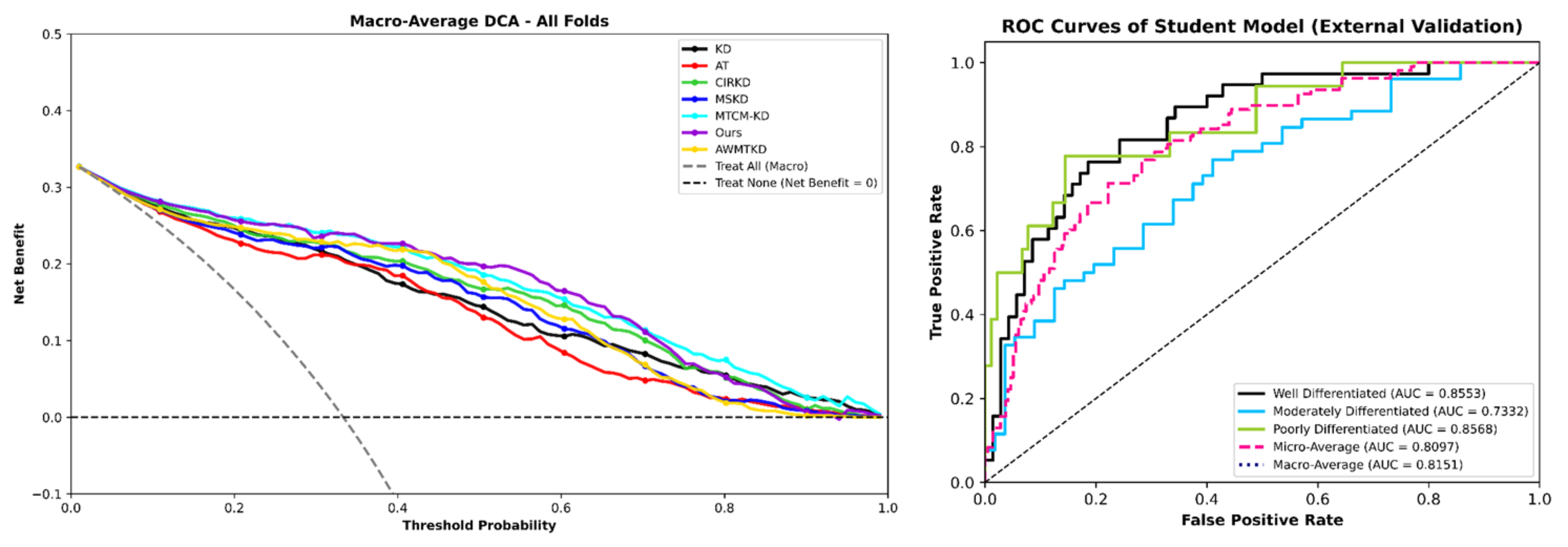}
    \caption{\textbf{Decision curve analysis (DCA, left) on internal validation and receiver operating characteristic (ROC, right) on external validation.} 
    The DCA curves indicate sustained net clinical benefit across decision thresholds, confirming the model’s practical reliability under degraded CT conditions, while the ROC curves demonstrate consistent discriminative capability across tumor grades.}
    
    \label{fig:roc_dca}
\end{figure}

To further assess clinical reliability, Fig.~\ref{fig:roc_dca} presents the receiver operating characteristic curve (ROC) under external validation and the decision curve analysis (DCA) based on internal validation. While the ROC curve measures the discriminative performance of the model, DCA provides a complementary evaluation by quantifying its clinical utility across a range of decision thresholds. Unlike traditional accuracy metrics, DCA considers the trade-off between true positives and false positives, offering insight into the potential benefit of applying the model in practice. In addition, we conducted DCA to compare different knowledge distillation strategies under internal validation, highlighting the clinical advantage of our proposed approach. In our study, the proposed model achieved strong discriminative performance in external testing (AUC~$> 0.76$) and demonstrated consistent positive clinical value in internal validation, supporting its robustness under real-world, low-quality imaging conditions.

\begin{figure}[t]
\centering
\includegraphics[width=\columnwidth]{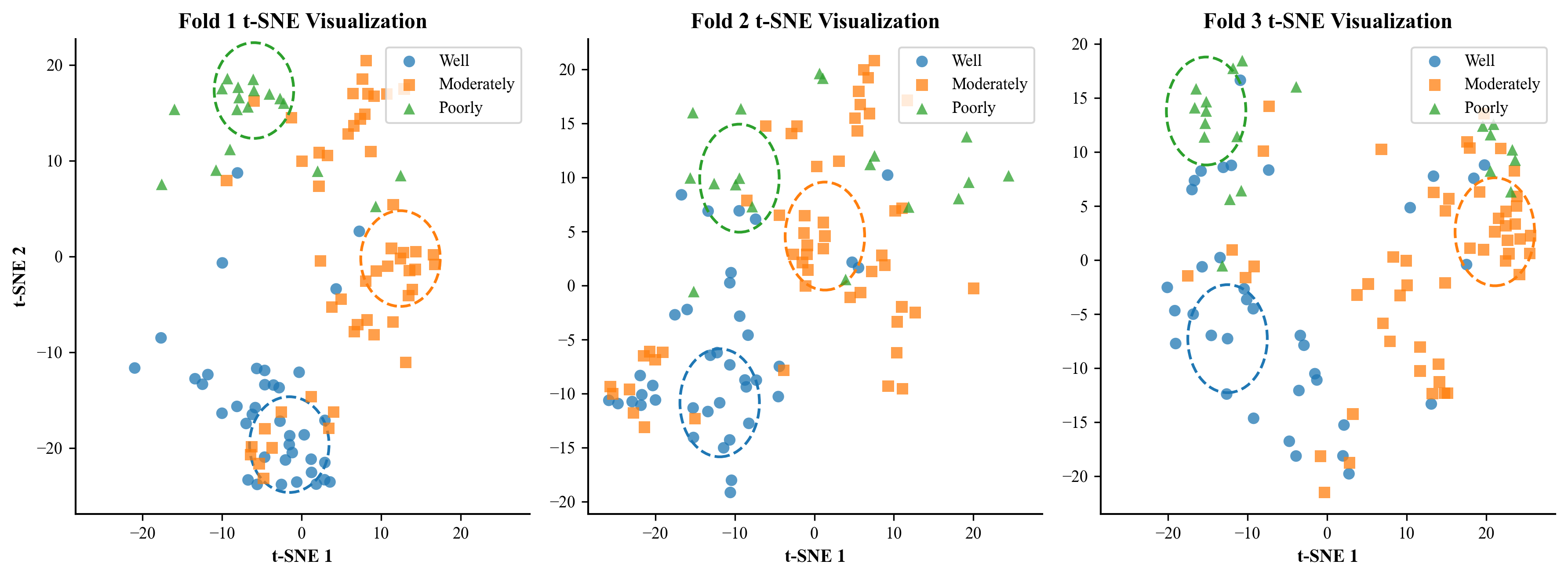}
\caption{\textbf{t-SNE projection of student logits on external test set.}  
Each subfigure corresponds to one cross-validation fold. The model trained with logits-only supervision yields incomplete separation of tumor grades, with particularly blurred margins between moderate and other differentiation levels. These findings highlight the insufficiency of global prediction signals in capturing nuanced class boundaries.}

\label{fig:SNE_high}
\end{figure}

\begin{figure}[t]
\centering
\includegraphics[width=\columnwidth]{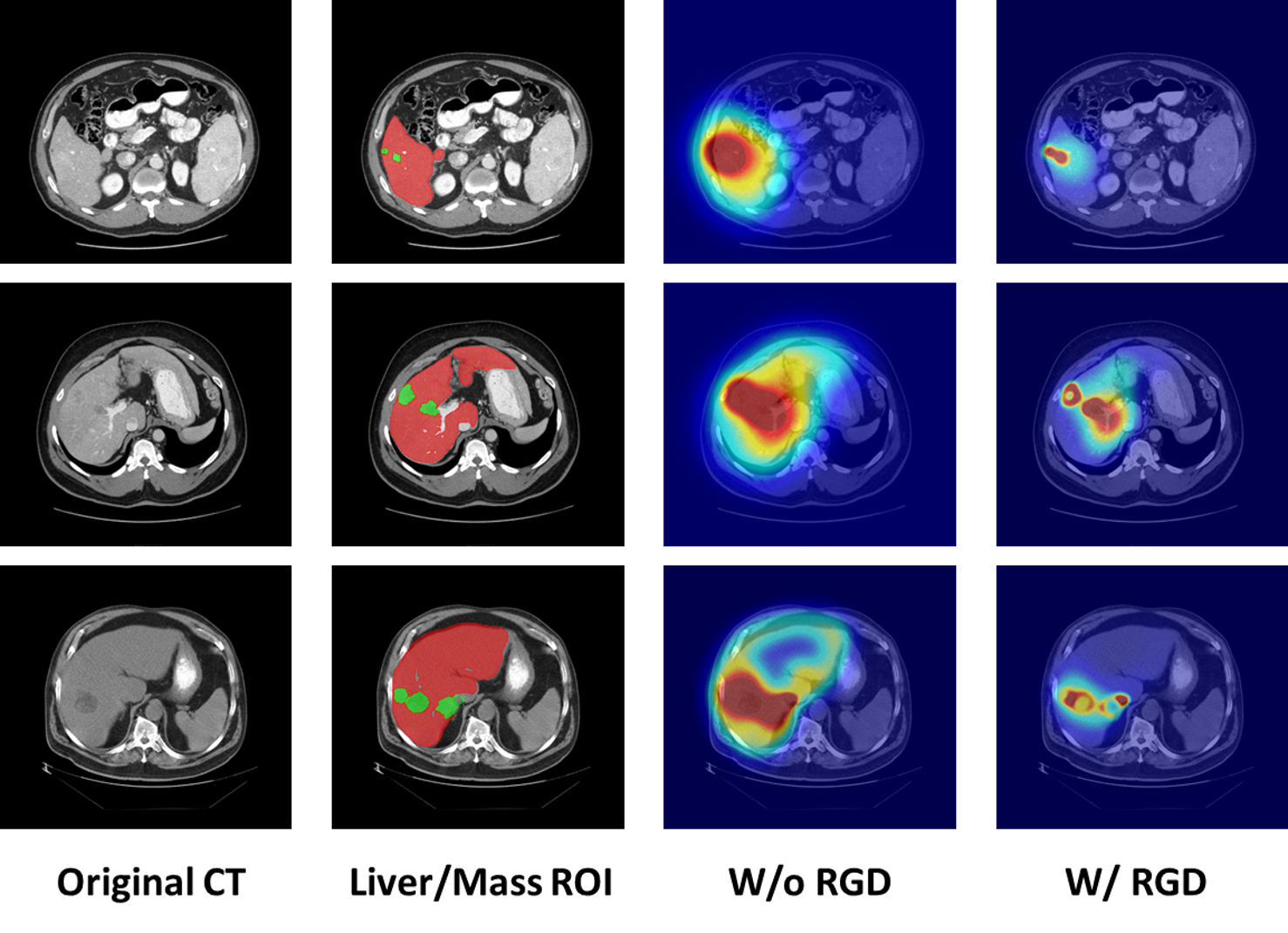}
\caption{\textbf{Topology-aware region graph visualization on external cases.}  
Each row depicts one test subject from the external cohort. From left to right: original CT image, anatomical ROIs (liver in red, tumors in green), attention maps from the student trained without region graph distillation (RGD), and the student trained with RGD. }

\label{fig:topo_visual}
\end{figure}

\subsection{Topology-Aware Representation and Interpretability}

To examine whether conventional knowledge distillation offers sufficient semantic supervision, we first visualize the t-SNE projection of final-layer logits from a student model trained with logits-only guidance. As shown in Fig.~\ref{fig:SNE_high}, although coarse class separation is evident, the embedding space still reveals substantial overlap between adjacent tumor grades, particularly within the moderately differentiated group. This observation suggests that global predictions alone may be insufficient to capture the complex decision boundaries characteristic of HCC.

To address this limitation, we introduce region graph distillation (RGD), which incorporates structural priors through anatomically grounded topological graphs. Each graph is constructed from liver and tumor segmentation masks, where each node represents a semantically distinct anatomical region, specifically the liver or an individual tumor lesion. These graphs encode both spatial relationships and lesion multiplicity, providing a structured basis for knowledge distillation.

The resulting node-level attention or embeddings can be projected back to the original 3D spacing, enabling interpretable visualization of the model's decision process. As illustrated in Fig.~\ref{fig:topo_visual}, the distilled student model generates activation patterns that reflect anatomical structure, capturing both central and dispersed lesions. This topology-aware interpretability, absent in standard distillation methods such as AT~\cite{zagoruyko2016paying}, MSKD~\cite{mskd}, supports more reliable and transparent deployment in clinical settings.

\section{Conclusion and Future Work}

We proposed a dual-teacher knowledge distillation framework for HCC tumor grade classification, integrating functional and dose-aware supervision through region graph distillation and attention-guided modeling. To the best of our knowledge, this is the first topological graph-based distillation framework to provide region-level visualization aligned with semantic anatomy, offering novel interpretability at both embedding and spatial levels. The distilled student model demonstrates robust performance across modality dropout and simulated dose degradation, while offering topology-aware interpretability grounded in anatomical structures. This work primarily focuses on high-level logits and graph-based distillation. 

Future extensions will explore feature-level transfer techniques under both homogeneous and heterogeneous settings, aiming to improve fine-grained representation learning. We also plan to evaluate the framework on other biomedical tasks, such as lesion segmentation and outcome prediction, to assess its generalizability across modalities and clinical objectives.

\section*{Ethics Statement}
This study protocol was reviewed and approved by the Ethics Committee of The Hong Kong Polytechnic University, approval number [HSEARS20241206005].

\section*{Acknowledgement}

This work was supported by an internal grant from The Hong Kong Polytechnic University (Project No. P0051278, Jung Sun Yoo) and the General Research Fund (Project No. PolyU 15101422, Jung Sun Yoo) from the Research Grants Council of the Hong Kong Special Administrative Region, China.

\bibliography{ref}  
\bibliographystyle{IEEEtran}  

\end{document}